\documentclass[%
 aip,
 amsmath,amssymb,
 reprint,%
]{revtex4-1}

\usepackage[utf8]{inputenc}
\usepackage{physics}
\usepackage{siunitx}
\usepackage{bm}
\usepackage[utf8]{inputenc}
\usepackage[T1]{fontenc}
\usepackage{mathptmx}
\usepackage{etoolbox}
\usepackage{hyperref}
\usepackage{svg}
\usepackage{graphicx}
\usepackage{color}

\makeatother
\begin{document}

\title{A motorized rotation mount for fast and reproducible optical polarization control}
\author{Adarsh P. Raghuram}
\author{Jonathan M. Mortlock}%
\author{Sarah L. Bromley}%
\author{Simon L. Cornish}%

\affiliation{$^1$Department of Physics, Durham University, South Road, Durham DH1 3LE, United Kingdom}%

\begin{abstract}
We present a simple motorised rotation mount for a half-wave plate that can be used to rapidly change the polarization of light. We use the device to switch a high power laser beam between different optical dipole traps in an ultracold atom experiment. The device uses a stepper motor with a hollow shaft, which allows a beam to propagate along the axis of the motor shaft, minimising inertia and mechanical complexity. A simple machined adapter is used to mount the wave plate. We characterise the performance of the device, focusing on its capability to switch a beam between the output ports of a polarizing beamsplitter cube. We demonstrate a switching time of 15.9(3)~ms, limited by the torque of the motor. The mount has a reaction time of 0.52(3)~ms, and a rotational resolution of 0.45~degrees. The rotation is highly reproducible, with the stepper motor not missing a step in 2000 repeated tests over 11 hours.

\end{abstract}
\maketitle

\section{Introduction}

Dynamic control of the polarization of light is a common requirement in many optical experiments. The rotation of a polarizer or a wave plate using a purpose-built rotation mount is a common method of polarization control. Desirable characteristics of a motorized rotation mount include high resolution, short rotation times, high repeatability, smooth rotation profiles and simple control. Common designs use servo motors and stepper motors \cite{fueten1997203,Shelton,Rakonjac_2013,Nilsson2021}, where the rotation of the shaft is transmitted to a receptacle for the optic using a series of gears \cite{Rakonjac_2013} or a belt \cite{Nilsson2021}. Commercial rotation mounts tend to prioritise resolution over speed and typically take over 100~ms to rotate 45 degrees.

Polarization control is frequently used to switch the power of a laser beam between different optical paths \cite{Nagano:18,Dimitrova2020,Kwon_2022}. In experiments involving ultracold matter, where high-power laser beams are commonly used to trap atoms \cite{Grimm}, it is often desirable to be able to divert the output of a single high-power laser between multiple traps during the experimental sequence.
For a linearly polarized beam of power $P_\text{0}$ incident on an ideal polarizer, the power of the transmitted beam $P_\textrm{T}$ is given by Malus' law, $P_\textrm{T}=P_\text{0}~\text{cos}^2(\theta)$.
Here, $\theta$ is the angle between the polarization of the incident beam and the axis of the polarizer. It follows that a 45-degree rotation of a half-wave plate (HWP) can be used to completely switch a beam between the output ports of a polarizing beam splitter (PBS) cube. 


This work is motivated by the need to switch a 30~W laser beam between two optical dipole traps in less than 50~ms, whilst avoiding losses associated with the limited diffraction efficiency of acousto-optic modulators.  Our solution is designed around a stepper motor with a hollow shaft, similar to the device used in \cite{Miranda}. Such devices are designed to allow for use of custom motor shafts or the passage of wiring through the motor. In our case, we transmit a laser beam through the hollow shaft of the motor and attach a HWP directly to the end of the shaft. This design has the advantage that the axis of rotation of the motor passes through the centre of the HWP, reducing the moment of inertia and enabling fast switching times. Below, we report details of the design and performance of our rotation mount. We show that it is able to reproducibly switch light between the two output ports of a PBS with switching times up to 15.9(3)\,ms.


\section{The Rotation Mount}
\begin{figure}[h!]
\includegraphics[width=8.5cm, trim=0cm 5cm 1.5cm 5cm,clip]{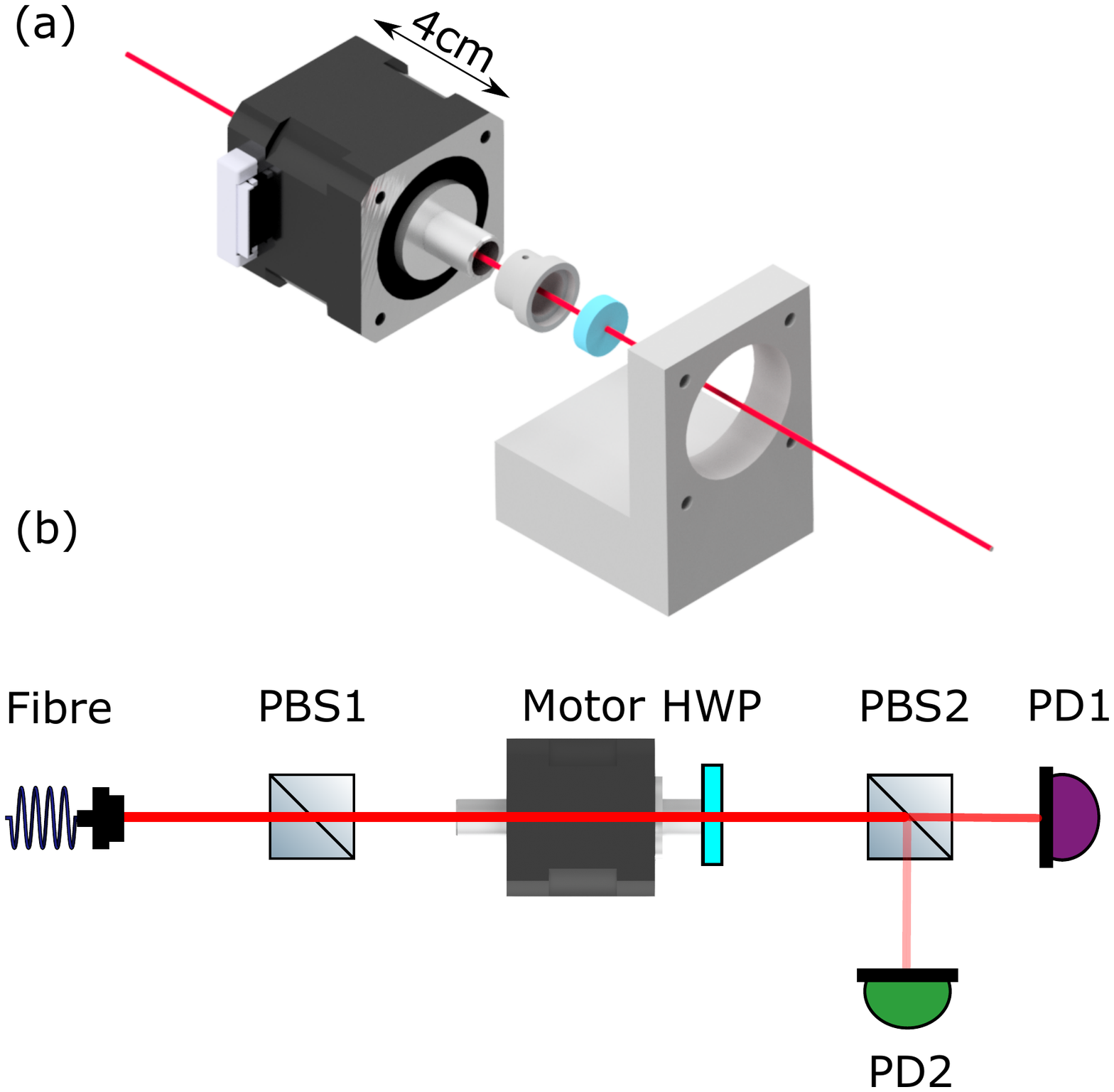}
\caption{(a) The hollow stepper motor along with the adapter for a half-inch diameter waveplate and an L-shaped mounting bracket. The motor shaft is hollow, with an inner diameter of 8~mm which allows a laser beam to pass through it as shown. (b) The setup used to test the rotation mount. The first polarizing beam splitter (PBS) cube sets a well-defined polarization into the rotation mount. By measuring the power out of the two ports of a second PBS with photodiodes (PD), we can extract the angle of the half-wave plate (HWP), independent of power fluctuations of the beam. 
\label{fig:setup}}
\end{figure}

The rotation mount uses a stepper motor with a hollow shaft, with adapters to attach the waveplate to the motor shaft, and the motor to an optical breadboard (Fig. \ref{fig:setup} (a)). We use the SCA4218M1804-L NEMA-17 hollow stepper motor \cite{WinNT}. The NEMA-17 model was selected for its balance of torque and mechanical size. The motor shaft has an inner diameter of 8 mm. The half-inch diameter waveplate is attached to the motor shaft using a compact adapter, minimising unnecessary inertia and hence maximising the rotary acceleration of the mount. In a stepper motor motion is discretized into well-defined steps, which allows for reliable open-loop control. The full step size of this motor is 1.8 degrees, however stepper motors can also be run under a micro-stepping configuration, which reduces the step size, allowing for smoother rotation profiles and higher resolution \cite{Acarnley}. We operate under half-stepping and quarter-stepping, which give step sizes of 0.9 and 0.45 degrees, respectively. Further divisions are possible and would allow even more precise motion, but this was not required for our application. 

The electronics required to operate the motor are readily available due to the popularity of stepper motors in hobbyist 3D printing. An Arduino UNO microcontroller is used to interface with a motor driver chip which provides the current waveform for the motor. We tested two driver chips, the DRV8825 and the TMC2208, and found vastly reduced mechanical vibrations with the TMC2208.  

\subsection{The test setup and general operation} 

For testing the rotation mount, we use a setup where the mount diverts an incident beam between the two output ports of a PBS and onto two photodiodes (Fig. \ref{fig:setup}(b)). The orientation of the HWP sets the polarization of the light into PBS2, and hence the fraction of the beam that reaches each photodiode. To ensure a well-defined linear polarisation into the rotation mount we use another polariser, PBS1.

We operate the rotation mount in two different configurations - the testing configuration and the switching configuration. In the testing configuration (the grey area of Fig. \ref{fig:time}(a)), the mount rotates between half maxima of the power to maximise angular sensitivity at the start and end of a 45 degree rotation.  In our ultracold matter experiments, we use the switching configuration (the pink area of Fig. \ref{fig:time}(a)), where the beam is switched between two different paths over the course of a rotation. We define the switching time as the time taken for a 45~degree rotation (a switch). We can use the relative power of the transmitted beam to calculate the angular orientation of the HWP using Malus' law. The performance of the rotation mount is assessed using the testing configuration.

For optimal performance of the stepper motor, we use a constant acceleration ramp, where we accelerate the motor to a maximum velocity and then decelerate to rest over the course of a switch. We use quarter micro-stepping to improve the resolution and reduce vibrations. The required waveform is generated by the Arduino Accelstepper library \cite{Accelstepper}. With a linear velocity ramp, the time between the m$^{\textrm{th}}$ and (m+1)$^{\textrm{th}}$ pulse into the stepper motor is given by \cite{Kenjo},
\begin{equation}\label{eqn:tm}
    t_\text{m} = 2\bigg(\sqrt{\frac{m+1}{a}} - \sqrt{\frac{m}{a}} \bigg),
\end{equation}

\noindent where $a$ is the acceleration of the motor in steps $\text{s}^{-2}$. Given that the first pulse occurs at $t=0$, we consider the time for the motor to turn is the time taken for $d$ pulses, where $d$ is the number of steps required to make the rotation (rotation angle/step size).

\subsection{Control of Rotation Dynamics}
\begin{figure}[h!]
\includegraphics[width=8.5cm]{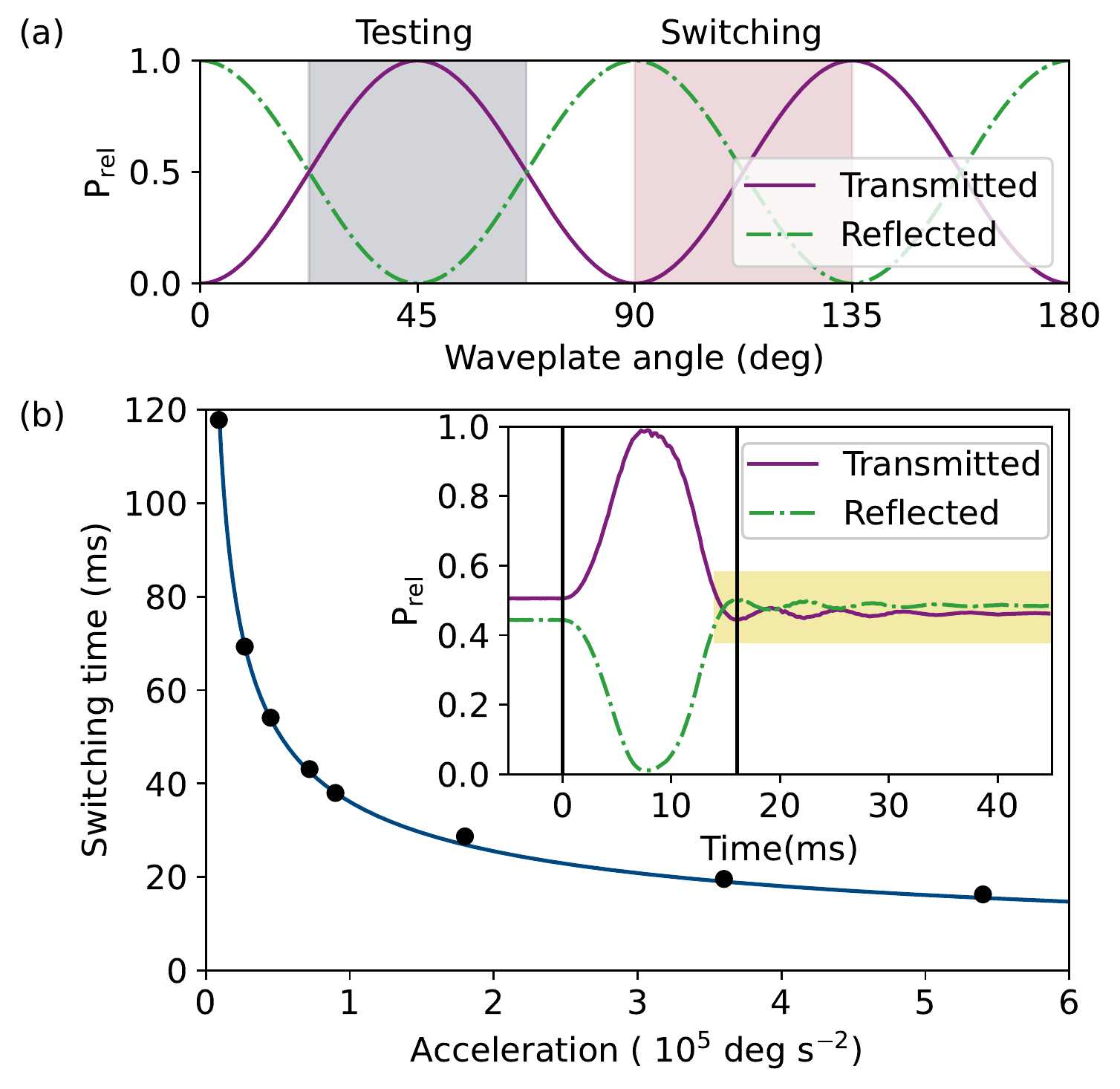}
\centering
\caption{\label{fig:time}(a) The predicted beam power as a function of waveplate angle. P$_\text{rel}$ refers to the power measured at a PD normalised to the full beam power. The grey and pink areas correspond to the testing and switching configurations, respectively. (b) Switching time vs. acceleration using half-stepping. The blue line is the predicted behaviour given in Eq.~\ref{eqn:tm}, and the black points are the measured times. To determine the time for 45-degree rotation we measure from the mount trigger signal to the first inflection point after the switch of the photodiode signal, as shown in the inset.  Each  point is the average of five measurements and was taken with the motor current limit set to 1.0 A. The typical uncertainty of the switching time is less than 0.4~ms.}
\end{figure}
The profile of a single switch in the testing configuration is shown in the inset of Fig. \ref{fig:time}(b). The switch is triggered by a pulse to the microcontroller. The time between the trigger pulse and the motor starting to move is the reaction time of the device. The rotation mount then rotates the HWP through 45 degrees, changing the power between the two beam paths. We observe a damped oscillation of the motor at the end of every switch, a known feature of stepper motors \cite{Acarnley}.

The reaction time of the motor is measured as 0.52(3)\,ms. We measure this as the time taken after the trigger pulse for the power transmitted through the PBS to change by 0.5\% of the initial power (the smallest value we can reliably resolve). At the half maxima, this corresponds to a rotation of 0.07\,degrees. 

Figure \ref{fig:time} (b) shows the switching time plotted as a function of acceleration alongside the expected behaviour from Equation 1. The switching time is measured as the time between the trigger pulse and the first inflection point after the switch (black lines in Fig. \ref{fig:time}(b) inset). It is the point where the motor is no longer moving in the direction it was during the switch.

 At the end of the motion there is an obvious oscillation of the shaft with an amplitude of between 0.4 and 0.7~degrees. When operating in the intended switching configuration, the effect of this oscillation on the power is greatly reduced, owing to the $\cos^2(\theta)$ dependence of Malus' law. This is seen in the data shown in the inset of Fig. \ref{fig:torque}(a).

\subsection{Maximum rotational acceleration.}
\begin{figure}

\includegraphics[width=8.5cm]{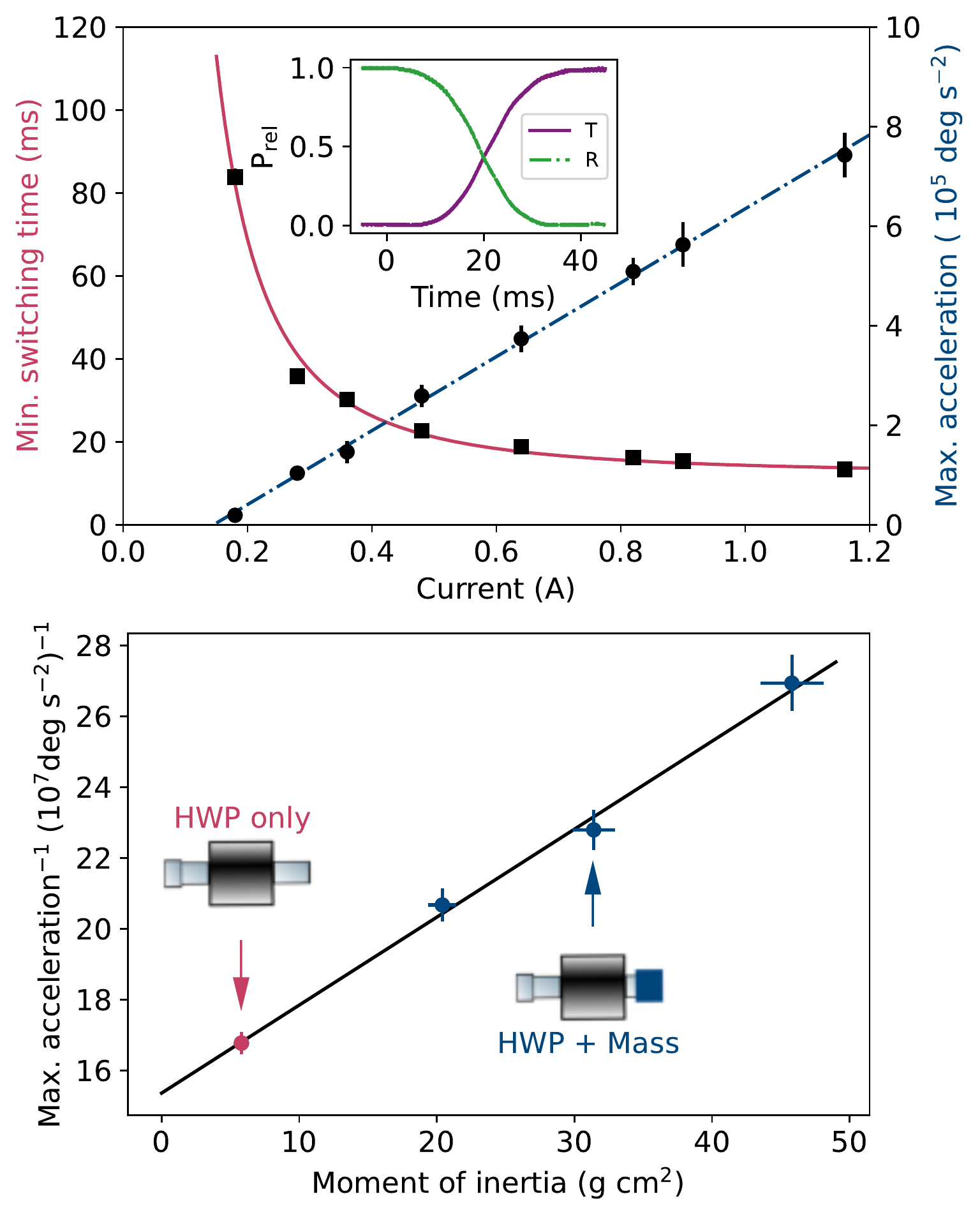}

\caption{\label{fig:torque} Switching time limited by torque of the motor. (a) Minimum switching time and maximum acceleration as functions of current. The maximum acceleration is directly proportional to the current, implying torque-limited behaviour. Inset is a typical switch in the switching configuration at 0.96~A, showing the diversion of power between the two paths. (b) The inverse of maximum acceleration as a function of the moment of inertia of load at a current of 0.96~A. With the waveplate fixed to one end, we attach hollow cylinders of varying moments of inertia to the motor shaft and measure the maximum acceleration.  The linear fit between acceleration and moment of inertia gives the torque as 0.069(3)~N\,m, and the moment of inertia of the shaft as 60(4)~g\,cm$^2$. The red point is the acceleration with only the waveplate attached}
\end{figure}

For a fast rotation mount, it is crucial to determine what limits the rotational acceleration. In Fig. \ref{fig:torque}, we show that the switching time is limited by the torque the motor can deliver. As shown in Fig. \ref{fig:torque} (a), we vary the current supplied to the motor, and observe the effect on the maximum acceleration (measured to 5\%) and hence the minimum switching time. The maximum torque output ($\tau$), and hence the maximum acceleration, of a stepper motor is related to the current into the motor winding ($I$), by a proportionality constant $K$, such that $(\tau = KI)$. This results in an inverse square relation between the switching time and the current. 
 
We further show that the torque output is what limits the switching time by attaching various masses with different moments of inertia to the shaft and observing their effect on the maximum acceleration at a fixed current of 0.96~A. As shown in Fig.~\ref{fig:torque}~(b), we find that the maximum acceleration is inversely proportional to the moment of inertia of the load. The maximum acceleration ($a_\text{max}$) is related to the moment of inertia of the load ($L_\text{load}$) as,

\begin{equation}\label{eq:amx}
    a_\text{max} = \frac{KI}{L_\text{shaft} + L_\text{load}}.
\end{equation}\\

\noindent Here, $L_\text{shaft}$ is the moment of inertia of the motor shaft itself, which acts as an absolute limit to the acceleration. $L_\text{load}$ includes the moment of inertia of the HWP and adapter, which we estimate to be 5.8(3)~g\,cm$^2$.  Fitting Eq.~\ref{eq:amx} to the results in Fig.~\ref{fig:torque}~(b) gives us the moment of inertia of the motor shaft as 60(4)~g\,cm$^2$ (in reasonable agreement with the manufacturer's value of 54~g\,cm$^2$), and the constant relating acceleration and torque as $K=0.072(3)$~N\,cm\,A$^\text{-1}$. At a current of 1.0 A, the maximum acceleration of the motor with the wave plate attached is 6.0(2)$\cross 10^5$~deg\,s$^\text{-2}$. The predicted maximum acceleration of the shaft with no load attached is $6.6(2)\times 10^5$~deg\,s$^\text{-2}$. We find that the motor works inconsistently above 1~A, and does not work at all above 1.3~A. So, we limit current to 1~A during operation. The quickest measured switching time at 0.96\,A is 15.9(3)~ms.



\subsection{Overshooting oscillations}

\begin{figure}
\includegraphics[width=8.5cm]{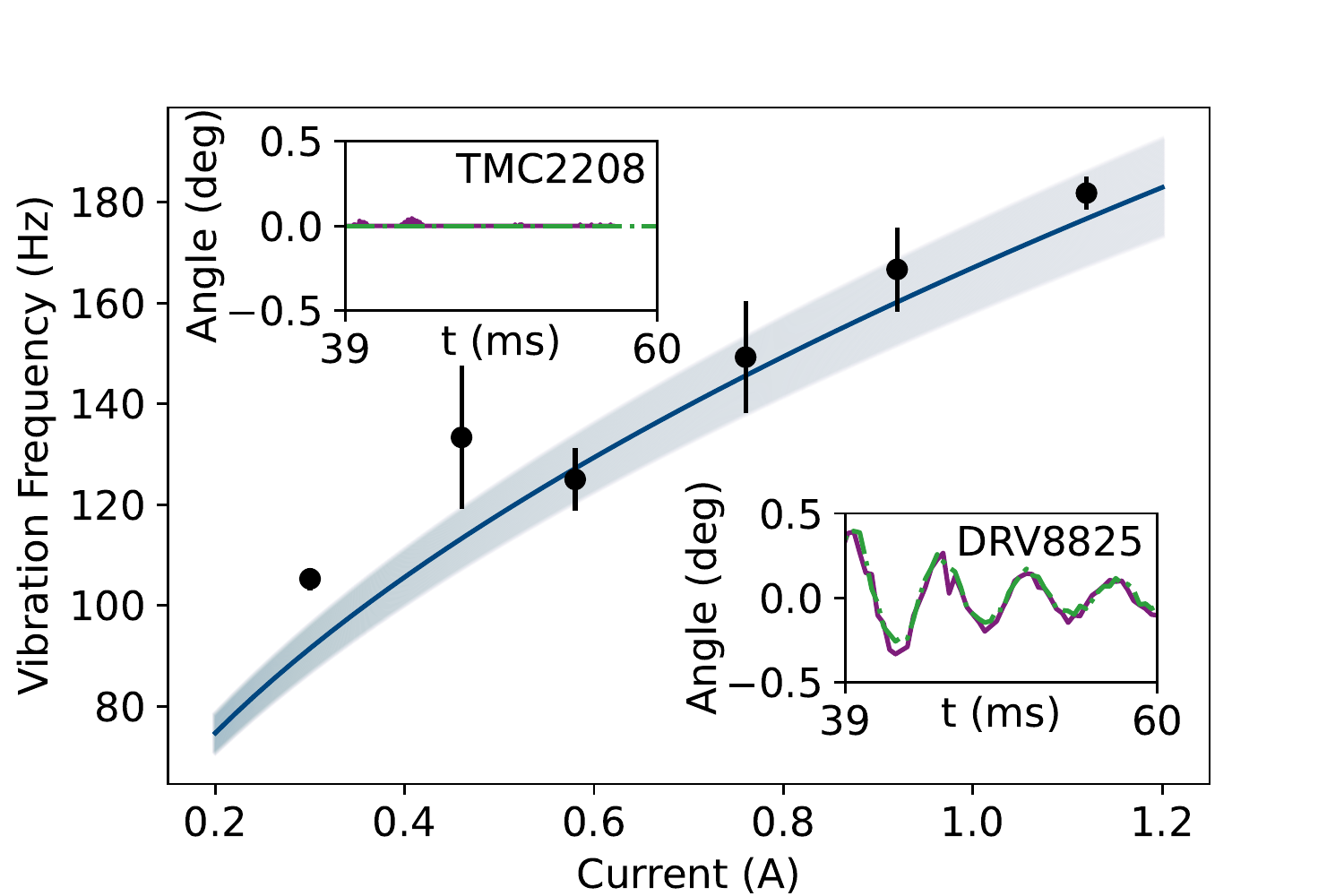}
\caption{Oscillation frequency as a function of the current for a switching time of 39~ms. After a switch, the stepper motor undergoes damped oscillation. The oscillation frequency is determined by the torque from the motor coils, \(\omega = \sqrt{KI/L}\) where the symbols are defined in Eq.~\ref{eq:amx}. The predicted frequency at different currents is shown by the blue region, with $\sqrt{K/L}$ taken as 167(9)~Hz/$\sqrt{A}$ calculated from values obtained earlier. The inset plots show the oscillation signal when using the DRV8825 and TMC2208 chips at a current of 0.92~A, and highlight the superior performance of the TMC2208.}
\label{fig:vibration}
\end{figure}

The stepper motor shows a damped oscillation after a switch, caused by the motor overshooting its intended position and undergoing a damped oscillation around the target position before coming to rest. We use two different drivers, the TMC2208 and the DRV8825, to investigate the effect of the driver chip on this oscillation. The oscillation frequency is proportional to the square root of the current (Fig.~\ref{fig:vibration}) as expected. Using the DRV8825, at switch times of 39~ms (acceleration $9\times10^4~\text{deg}~\text{s}^{-2}$), the maximum peak to peak amplitude of the oscillation corresponds to 0.4 degrees of angular vibration of the shaft, or 3\% of the peak signal in the testing configuration. In the switching configuration, this is not an issue, as the amplitude of the oscillation reduces to 0.03\% of the power. As shown in the inset, the TMC2208 chip is able to stop the occurrence of the oscillations by ramping off the current in the motor coils \cite{TMC} at switch times greater than 31.0(5)~ms. For switches quicker than 30.5(5) ms, the oscillations occur with similar amplitudes to that seen using the DRV8825 chip.

\subsection{Reliability and repeatability}
We tested the reliability and repeatability of the device by leaving it to run for 2000 switches over a period of 11 hours in a half-stepping configuration at a switching time of 39~ms. We set the motor to move forward 45 degrees every 20~s to check for any accumulation of error. A skipped micro-step during a switch would result in an offset in the position of the motor by 0.9 degrees (a half-step) for each subsequent switch. The motor should return to the same position every 8$^\textrm{th}$ switch.

We attach a mask to the rotation mount for this measurement. The mask is oriented such that one of its opaque regions with a sharp edge obscures half the beam at the start of a switch. We measure the power of the transmitted beam on a photodiode at the start of every full rotation of the mask, as well as the full power, midway through the switch when the entire beam is incident on the photodiode. We then calculate the fractional transmission when the beam is cut in half, effectively normalizing for drifts in laser power over the course of the measurement. A skipped micro-step during a switch would appear as a 3\% change in fractional transmission at the start of a switch. The standard deviation of the fractional transmission over the full period of measurement is 0.12\%, with the single largest deviation being 0.37\%. We conclude that the motor did not miss a step in the 11 hours of testing and is therefore reliable enough for open loop operation.\\


\section{Conclusion}
In conclusion, we have presented a motorised rotation mount that can be used to rapidly switch the polarization, and hence the path of a high power laser beam in under 16~ms. We have implemented the rotation mount in our ultracold atom experiment where a 45 degree rotation is used to re-purpose power from one laser to different beam paths. We are hence able to use the full power from one laser for different optical dipole traps at different times in our experimental sequence. Smaller rotations of the HWP can be used to control the power to each path, and hence the depths of the optical dipole traps. The ease of construction and operation of the device facilitate wide adoption in similar experiments where the switching of beam paths is a common problem. \\

\section*{Acknowledgements}
The authors thank Martin Miranda, whose work gave inspiration for this project and who contributed helpful discussion. This work was supported by UK Engineering and Physical Sciences Research Council (EPSRC) Grant EP/P01058X/1, UK Research and Innovation (UKRI) Frontier Research Grant EP/X023354/1 and Durham University. 

\section*{Author Declarations}
\noindent\textbf{Conflict of interest}\\

The authors have no conflicts to disclose.\\

\noindent\textbf{Author Contributions}

\textbf{A. P. Raghuram:} Conceptualization (supporting); Formal Analysis (lead); Investigation (lead); Methodology (equal); Writing - Original Draft Preparation (lead); Writing - Review \& Editing (equal) Visualization (lead) \textbf{J. M. Mortlock:} Conceptualization (lead); Formal Analysis (supporting); Investigation (supporting); Methodology (equal); Writing - Review \& Editing (equal); \textbf{S. L. Bromley:} Supervision (supporting); Writing - Review \& Editing (equal) \textbf{S. L. Cornish:} Funding Acquisition (lead); Supervision (lead); Writing - Review \& Editing (equal)

\section*{Data Availability Statement}
The data that support the findings of this study is published in the Durham University open data repository - DOI: http://doi.org/10.15128/r1n870zq866 \\

\bibliography{paper.bib}
\end{document}